\documentclass[aip,jcp,amsmath,amssymb,preprint]{revtex4-1}

\usepackage{graphicx}% Include figure files
\usepackage{dcolumn}% Align table columns on decimal point
\usepackage{bm}% bold math
\usepackage{comment}
\usepackage{xcolor}
\usepackage{soul}
\renewcommand{\vec}[1]{\mathbf{#1}}
\begin{document}

%\preprint{AIP/123-QED}

\title{Topological extension of the isomorph theory based on the Shannon entropy}
%Theoretical equivalence between thermodynamic entropy and Shannon entropy based on the isomorph theory}

\author{Tae Jun Yoon}
\affiliation{School of Chemical and Biological Engineering, Institute of Chemical Processes, Seoul National University, Seoul 08826, Republic of Korea}

\author{Min Young Ha}
\affiliation{School of Chemical and Biological Engineering, Institute of Chemical Processes, Seoul National University, Seoul 08826, Republic of Korea}

\author{Emanuel A. Lazar}
\affiliation{Department of Mathematics, Bar-Ilan University, Ramat Gan 5290002, Israel}

\author{Won Bo Lee}
\email{wblee@snu.ac.kr}
\affiliation{School of Chemical and Biological Engineering, Institute of Chemical Processes, Seoul National University, Seoul 08826, Republic of Korea} 

\author{Youn-Woo Lee}
\email{ywlee@snu.ac.kr}
\affiliation{School of Chemical and Biological Engineering, Institute of Chemical Processes, Seoul National University, Seoul 08826, Republic of Korea}
 
\date{\today}

\begin{abstract}
Isomorph theory {is one of the promising theories to understand the quasi-universal relationship between thermodynamic, dynamic and structural characteristics. Based on the hidden scale invariance of the inverse power law potentials, it rationalizes the excess entropy scaling law of dynamic properties. This work aims to show that this basic idea of isomorph theory can be extended by examining the microstructural features of the system. Using the topological framework in conjunction with the entropy calculation algorithm, we demonstrate that Voronoi entropy, a measure of the topological diversity of single atoms, provides a scaling law for the transport properties of soft-sphere fluids, which is comparable to the frequently used excess entropy scaling. By examining the relationship between the Voronoi entropy and the solid-like fraction of simple fluids,} we suggest that the Frenkel line, a rigid-nonrigid crossover line, {be} a topological isomorphic line where the scaling relation qualitatively changes.
\end{abstract}

\maketitle

\section{Introduction}

The dynamic behavior of particles in liquids and high-pressure supercritical fluids is largely influenced by the relative local configurations of their neighbors. This strong particle-particle correlation implies that the thermodynamic, dynamic, and structural characteristics of dense fluids are intimately linked with each other. Hence, it is no wonder that a considerable amount of studies were devoted to developing the theory of the liquid state \cite{eyring1937theory,hansen1990theory}. One of the wonderful aspects of the liquid state theory is its simplicity based on the hard-sphere paradigm \cite{dyre2016simple}. The hard-sphere paradigm assumes that the repulsive part of interatomic interaction dominates the behavior of the liquid state. Based on the hard-sphere paradigm, the perturbation theory \cite{verlet1972equilibrium,barker1967perturbation} has been advanced to understand the thermodynamic behavior of dense fluid systems based on the pair correlation function and the hard-sphere potential as a reference system \cite{carnahan1969equation}.

Liquid state theory has also been used to relate the thermodynamic properties to transport properties. Rosenfeld \cite{rosenfeld1977relation} and Dzugutov \cite{dzugutov1996universal} proposed the scaling relation that connects the thermodynamic excess entropy ($S_{\rm{exc}}$) and the scaled transport properties of dense fluids. Rosenfeld discovered that the scaled diffusivities of simple liquids modeled with different interatomic potentials collapse to a single line as a function of the thermodynamic excess entropy. Based on these scaling laws, the two-body excess entropy ($S_{2}$), which can be directly obtained based on the pair correlation function, has been frequently used when scaling the dynamic properties based on the structural characteristics. The scaling relation provided by $S_{2}$ was fairly good for simple fluid models \cite{ding2015equilibrium} {although the contribution of $S_{2}$ on the total excess entropy varies depending on thermodynamic conditions \cite{baranyai1990three,chopra2010use,hoyt2000test,jakse2016excess,yokoyama2002excess}.}

  {In a more recent approach, Dyre and his coworkers proposed the isomorph theory to understand the relationship between thermodynamic, dynamic and structural characteristics of simple fluid systems in an integrated manner.} 
  {
    In the isomorph theory, the two \textit{state points} are defined to be isomorphic if one can find pairs of scaled configurations that have the same canonical probability~\cite{dyre2013isomorphs}. Let two configurations $\vec{R}_A$ and $\vec{R}_B$ ($\vec{R}_{i}\equiv(\vec{r}_i^{1},\vec{r}_i^{2},\ldots,\vec{r}_i^{N})$) sampled from two thermodynamic state points $(\rho_A, T_A)$ and $(\rho_B, T_B)$, respectively, have the same reduced densities, i.e. $\rho_A^{1/3} \vec{R}_A = \rho_B^{1/3} \vec{R}_B$. The two state points are isomorphic if one has
}
\begin{equation}
    {
        \exp\left( -\frac{U(\vec{R}_A)}{k T_A} \right) = \mathcal{C}_{AB}\exp\left( -\frac{U(\vec{R}_B)}{k T_B} \right).
    }
\end{equation}

\noindent {
    where $\mathcal{C}_{AB}$ is a configuration-independent constant. Gnan et al. have shown that the condition of having a good isomorph is equivalent to having a strong correlation between fluctuations of virial and potential-energy, which they term Roskilde-simple (R-simple) liquid~\cite{gnan2009pressure}.
}
The R-simple liquid is defined as fluid models {of which the virial potential-energy correlation ($\mathcal{R}$) is higher than $0.9$. Here, the correlation coefficient $\mathcal{R}$ is defined as:} 
\begin{equation}
    {\mathcal{R}=\frac{\langle\Delta{W}\Delta{U}\rangle}{\sqrt{\langle(\Delta{W})^2\rangle\langle(\Delta{U})^2\rangle}},}
    \label{eqn:virial-energy}
\end{equation}
\noindent {where $W$ is the virial ($W=pV-Nk_{B}T$), $U$ is the potential-energy, and $\Delta{A}$ is a deviation ($\Delta{A}=\langle{A}\rangle-A$).} {Schr{\o}der and Dyre demonstrated that the following conditional proposition for two system $A$ and $B$ is exact when the correlation coefficient $\mathcal{R}$ is unity \cite{schroder2014simplicity}:}

\begin{equation}
    {\rho_{A}^{1/3}\vec{R}_{A}=\rho_{B}^{1/3}\vec{R}_{B} \Rightarrow{S_{exc}(\vec{R}_{A})=S_{exc}(\vec{R}_{B})}.}
    \label{eqn:isomorph-definition}
\end{equation}

\noindent {In Eqn. (\ref{eqn:isomorph-definition}), $\rho_{i}$ and $S_{exc}(\vec{R_{i}})$ are the bulk density and the excess entropy of the system $i$.} When two systems satisfy the antecedent, they are regarded to be isomorphic to each other. They showed that the Newton's second law of motion in reduced units is invariant for the isomorphic states:
\begin{equation}
    {\rho_{A}^{1/3}\vec{R}_{A}=\rho_{B}^{1/3}\vec{R}_{B} \Rightarrow{\vec{\tilde{F}}_{A}=\vec{\tilde{F}}}_{B},}
    \label{eqn:reduced-newton}
\end{equation}
\noindent {where $\vec{\tilde{F}}$ is dimensionless force vector ($\vec{\tilde{F}}\equiv\rho^{-1/3}\vec{F}/k_{B}T$). This result explains why the excess entropy scaling law holds for simple fluids.} In subsequent articles, they successfully showed that the isomorph theory works as a ``good approximation'' to different types of potentials \cite{bacher2014explaining} including Lennard-Jones~\cite{bohling2012scaling}, Yukawa~\cite{veldhorst2015invariants} and exponential pair potentials \cite{bacher2014mother}. Some known exceptions, that do not follow the isomorph theory and the excess entropy scaling law, are the potentials with thermodynamic anomalies \cite{fomin2010breakdown}.

Despite this success of the isomorph theory, it should be noted that a direct connection between the structural definition of isomorphic states and the dynamics scaling was not given yet. Unlike crystalline systems, it is extremely difficult to discover two liquid configurations that exactly satisfy the antecedent of Eqn. (\ref{eqn:isomorph-definition}). A pair correlation function, $g(r)$, has been frequently used as an indicator of the antecedent, but {the details of the local configuration cannot be inferred from the pair correlation function.} Moreover, the two-body excess entropy directly calculated from the pair correlation function cannot work as a robust parameter because its proportion varies depending on thermodynamic conditions \cite{baranyai1990three,chopra2010use,hoyt2000test,jakse2016excess,yokoyama2002excess}. 

  {We note that the antecedent of Eqn. (\ref{eqn:isomorph-definition}) can be reformulated from the atomistic point of view. The scale invariance hypothesized in Eqn.~(\ref{eqn:isomorph-definition}) can be expressed as follows.} Let $\vec{\xi}_{i}^{j}$ be a position vector of the neighbor atom $j$ relative to the atom $i$ ($\vec{\xi}_{i}^{j}=\vec{r}^{j}-\vec{r}^{i}$), and $\vec{\Xi}_{i}\equiv(\vec{\xi}_{i}^{1},\vec{\xi}_{i}^{2},  {\vec{\xi}_{i}^{3}},\cdots)$ a set of position vectors. Then, two local configurations of particles $a$ and $b$ are isomorphic to each other if the following condition is satisfied:
\begin{equation}
    \rho_{a,l}^{1/3}\vec{\Xi}_{a}=\rho_{b,l}^{1/3}\vec{\Xi}_{b}\Rightarrow{S_{exc}(\vec{R}_{A})=S_{exc}(\vec{R}_{B}),}
    \label{eqn:atomistic-isomorph-definition}
\end{equation}
\noindent where $\rho_{i,l}$ the local density of the atom $i$  {, and $\Xi_{i}$ contains all particles of the system except the particle $i$. This microscopic definition itself does not provide any advantages over the macroscopic description, but this point of view can extend the definition of the isomorphic states in conjunction with the notion of the Gibbs entropy, which states that the system entropy is given by a distribution on the microstates ($S=-k_{B}\sum{p_{i}}\log{p_{i}}$ where $p_{i}$ is the probability of the microstate $i$). In a similar vein, if a relative configuration ($\rho_{i,l}^{1/3}\Xi_{i}$) is regarded as a \textit{microstate}, we can hypothesize that two configurations will have the same excess entropy if their distributions of the relative configurations are the same.}

  {
    This viewpoint is related to our works on dense supercritical fluids. We have characterized the local structure of an atom with respect to the topological type of its Voronoi polyhedron to develop a theory of structure-dynamics relationship~\cite{yoon2018topological} and a notion of quasi-universality among simple fluids~\cite{yoon2019topological}. In this work, we validate the idea of Eqn. (\ref{eqn:atomistic-isomorph-definition}) by defining the microstate of an atom as the topological type of its Voronoi polyhedron~\cite{lazar2015topological}, and estimating the excess entropy from the diversity of this topological type in the given thermodynamic condition, where the local density $\rho_{i,l}$ in Eqn. (\ref{eqn:atomistic-isomorph-definition}) is given as the inverse volume of the Voronoi polyhedron. Then, the classical notion about the equivalence of the Shannon entropy and the thermodynamic entropy is exploited to define Voronoi entropy based on the topological types observed in the system. By comparing the scaling results of the repulsive $n-6$ models obtained from the thermodynamic excess entropy and the Shannon entropy, we not only demonstrate that the Voronoi entropy works as a good measure to scale the dynamic properties, but also test the equivalence of the Shannon excess entropy and the thermodynamic excess entropy} Lastly, we show that the Frenkel line, a rigid-nonrigid transition line {which can be understood with respect to the percolation of rigid microstates}~\cite{yoon2018two,yoon2018topological}, can be regarded as a limit of {applicability of} the exponential scaling relation.

\section{Methods}

\subsection{Molecular Dynamics (MD) simulations}
We perform the NVT simulations \cite{plimpton1995fast} of the soft-sphere fluids of which the interatomic potentials are modeled with the repulsive $n-6$ potential. The repulsive $n-6$ potential is given as follows.

\begin{equation}
    \phi(r)=
    \begin{cases}
      \phi_M(r)-\phi_M(r_{\rm{cut}}){\qquad}&r{\leq}r_{cut}\\
      0{\qquad}&r{\geq}r_{cut}\\
    \end{cases}
	\label{eqn:repulsive n-6}
\end{equation}
Here, $\phi_{M}(r)$ is the Mie $n$-6 potential, which is given in Eqn.~(\ref{eqn:Mie potential}).

\begin{equation}
    {\phi_M(r)=\left[\frac{n}{n-6}\right]\left(\frac{n}{6}\right)^{n/(n-6)}\epsilon\left[\left(\frac{\sigma}{r}\right)^{n}-\left(\frac{\sigma}{r}\right)^{6}\right]}
    \label{eqn:Mie potential}
\end{equation}
The potential is shifted and truncated at $r_{\rm{cut}}=(n/6)^{1/(n-6)}\sigma$. The size parameter $\sigma$ of argon is used for all potentials ($\sigma=3.405$\AA). The energy parameter $\epsilon$ is changed so that the coefficient $C_{n}\epsilon$ becomes equal to that of the LJ potential ($C_{n}\epsilon=4\epsilon_{Ar}$) where $\epsilon_{Ar}$ is the energy parameter of argon ($\epsilon_{Ar}/k_{B}=119.8 K$). The simulation temperatures are $T=318.29, 636.57, 954.86, 1273.1$ and $1591.4$ K. The repulsive exponents are $n=8-24$. For all simulations, the timestep is 2 fs. To obtain the trajectories for calculating the Shannon entropy, and the thermodynamic and transport properties, the systems are equilibrated for 100,000 steps. The details of the production run are given in the following subsections.

  {\subsection{Evaluation of the virial potential-energy correlation}}
  {Virial potential-energy correlation is evaluated for all thermodynamic conditions as follows. In the production run ($5,000,000$ steps), the instantaneous virial ($W$) and the potential-energy ($U$) are collected every ten steps. Then, the correlation coefficients $\mathcal{R}$ are evaluated using Eqn. (\ref{eqn:virial-energy}). As shown in the Supplementary Information, the correlation coefficients $\mathcal{R}$ are always higher than $0.98$ at all conditions. Hence, all repulsive $n$-6 fluids dealt with in this work are R-simple.}

\subsection{Topological framework for local structure analysis}

The topological framework for local structure analysis proposed by Lazar et al.~\cite{lazar2015topological} describes the arrangement of neighbors surrounding a central particle via the Voronoi tessellation, the partitioning of space into regions, each of which consist of all points closer to a given particle than to any other. The topology of a Voronoi cell can be described by enumerating the number of edges of each of its faces.  Although this description provides more information than a mere count of faces, it does not completely describe how a particle's neighbors are arranged relative to the central particle and to one another.  A more refined description of the Voronoi cell, and thereby of the arrangement of neighbors, is provided by the isomorphism class of its edge graph \cite{lazar2015topological}, which identifies two Voronoi cells as the same if pairs of faces are adjacent in one Voronoi cell if and only if corresponding faces in the other are also adjacent.  This connectivity information can be encoded as a series of integers called the Weinberg vector \cite{lazar2012complete}, {which is obtained from a graph-tracing algorithm initially developed to check whether two planar graphs are isomorphic \cite{weinberg1966simple}.} Hence, the Weinberg vector can be viewed as a `name' of the topological type of a Voronoi cell. We use the {\it VoroTop} library \cite{lazar2017vorotop} to gather the statistical data of the topological types discovered in the configurations.

\subsection{\label{sec:fl}Characterization of the Frenkel line}
Rosenfeld et al.\cite{rosenfeld1977relation} noted that there are two regions where the dependence of transport properties on the thermodynamic excess entropy are qualitatively different \cite{mittal2007relationships}. In the low-density (low excess entropy) region, the diffusivity shows a power-law dependence. When the density is high, it shows an exponential dependence on the excess entropy in the high-density region. This qualitative change of dynamics can also be observed in Monte Carlo simulations. Nezhad and Deiters \cite{nezhad2017estimation} recently discovered that the excess entropy is an approximately linear function of the reciprocal mean Monte Carlo displacement parameter at high density. Provided that the Monte Carlo displacement parameter is proportional to the diffusivity, this finding indicates that the collective particle dynamics changes depending on the bulk density of a system.

This qualitative change of the transport properties would be related to the Frenkel line proposed by Brazhkin et al. \cite{brazhkin2012two} They proposed that the Frenkel line of the hard-sphere fluid corresponds to the crossover density at which the transport properties show a qualitatively different dependence on the bulk density \cite{brazhkin2018liquid}. In recent work, we demonstrated that this conjecture is quite reasonable based on the topological framework \cite{yoon2018topological} and the two-phase thermodynamics (2PT) model \cite{yoon2018two}. In addition, we recently found that the percolation behavior of solid-like structures of different repulsive $n-6$ fluids collapses to a single line when the fraction of solid-like molecules ($\Pi_{\rm{solid}}$) was used as an order parameter \cite{yoon2019topological}. Hence, we validate this idea that the Frenkel line may be a good candidate to demarcate the fluid region depending on the behavior of the transport properties as proposed by Rosenfeld et al.

To locate the dynamic crossover conditions, we use the topological classification method proposed in our earlier works \cite{yoon2018topological,yoon2019topological}. In this method, the topological types of two dynamic limits of the fluid phase including the ideal gas and the maximally random jammed state are used to classify a molecule as either gas-like (diffusive) or solid-like (oscillatory). If a topological type of an atom discovered in a configuration has a higher likelihood to be observed in ideal gas, it is classified as gas-like. Otherwise, it is classified as solid-like. A weighted mean-field approximation is then applied to this initial classification result to remove the influence of fluctuation. From the finite-size scaling analysis on percolation behaviors, we showed that the Frenkel line can be defined as the thermodynamic states where the fraction of solid-like molecules ($\Pi_{\rm{solid}}$) reaches {the percolation threshold,} $\Pi_{\rm{solid}}=0.1159\pm0.0081$ \cite{yoon2018topological}. In this work, we apply the same procedure and the percolation criterion to determine the Frenkel line of the soft-sphere fluids (see Yoon et al. \cite{yoon2018topological,yoon2019topological} for further details of the algorithm).

  {\subsection{Reformulation of the isomorph definition based on the information theory}}
  {The limitation of the hypothesis provided in Eqn. (\ref{eqn:atomistic-isomorph-definition}) is that the definitions of $\rho_{i,l}$ and $\Xi_{i}$ are incomplete. As a first approximation, we introduce the topological framework proposed by Lazar et al \cite{lazar2015topological}. In the topological framework, the connectivity information of an atom with its nearest neighbors is understood based on the topological type of the Voronoi polyhedron. Since this topological information is invariant under the multiplication of coordinates by a constant, e.g. $\tilde{\vec{R}}_i = \rho_i^{1/3}\vec{R}_i$, the reduced coordinates of the nearest atoms surrounding two atoms are the same if the topological types of their Voronoi cells are identical. Hence, Eqn. (\ref{eqn:atomistic-isomorph-definition}) is approximated as:
\begin{equation}
    v_{a,l}^{-1/3}\vec{\Xi'}_{a}=v_{b,l}^{-1/3}\vec{\Xi'}_{b} 
    \label{eqn:new-isomorph-definition}
\end{equation}
where $v_{i,l}$ is the volume of the Voronoi polyhedron of the particle $i$ and $\vec{\Xi'}_{i}$ is a set of the relative coordinate vectors of the nearest neighbors of which the Voronoi polyhedron share a face with that of the central particle $i$. Since the forces exerted on the central atom by the nearest neighbors usually account for the majority of the total force, we expect that two atoms would have similar dynamic characteristics if the topological types of their Voronoi cells are identical to each other.} 

It is noteworthy that a similar extension of the isomorphism was proposed by Malins, Eggers, and Royall \cite{malins2013investigating}. They used the topological classification method proposed by Williams \cite{williams2007topological} to identify the bicapped square antiprism, which is a locally favored structure in glass formers.\\

\subsection{Voronoi entropy}
The diversity of the categorical distributions can be measured using the Shannon entropy \cite{jost2006entropy,masisi2008use,cover2012elements}. The Shannon entropy ($\mathcal{H}$) is obtained as:
\begin{equation}
    \mathcal{H}=-\sum_{i}p_{i}\log{p_{i}}
    \label{eqn:Shannon-entropy}
\end{equation}
where $p_{i}$ is the probability of finding a topological type $i$ in the system. The term Voronoi entropy was used by Peng, Li and Wang \cite{peng2013stress} by applying Eqn.~(\ref{eqn:Shannon-entropy}) to the distribution of the Voronoi types, which were classified based on their Voronoi indices. {It was also defined as the Shannon entropy of the distribution of the Voronoi types based on the number of edges \cite{bormashenko2019voronoi}.} {On the other hand}, we classify Voronoi cells based on the Weinberg vectors, a more refined descriptor than the Voronoi indices, {following the philosophy of the isomorph theory.}

When the probability that an event $i$ occurs ($p_{i}$) is known for all events, we can directly measure the Shannon entropy of a system. In real-world problems, however, {there are two bottlenecks to apply Eqn. (\ref{eqn:Shannon-entropy}) directly. First,} $p_{i}$ is only estimated based on observation of the limited samples drawn from a population. The Shannon entropy calculated based on this limited observation can be heavily biased by rare events. For systems in which the number of events is infinite (unbounded), therefore, the Shannon entropy is exactly calculated only when infinite data are available or an exact mathematical expression for all the $p_{i}$'s is given. For ideal gas, the number of topological types is infinite since the point particles can be randomly distributed in a system.  {Second, Eqn. (\ref{eqn:Shannon-entropy}) ignores the correlation between events. For ideal gas, this hypothesis holds since no interatomic interaction exists among the particles. That is, the topological type of an atom has little effect on how a neighbor atom is surrounded by its neighbors in the low-density regime. On the other hand, the probability of finding a topological type is largely influenced by the topological types of its neighbors in the crystalline state.} 

  {Several algorithms have been proposed to resolve the problem of infinite sample size by estimating the upper bound of the Shannon entropy ($\hat{\mathcal{H}}$) with an unknown or infinite number of samples \cite{archer2012bayesian,chao2003nonparametric,valiant2013estimating}.} This work uses the estimator named \textit{Unseen} designed by Valiant and Valiant \cite{valiant2013estimating}. This algorithm uses a fingerprint of a finite dataset (observed samples), a histogram of a histogram, to construct a plausible histogram of which the entropy and other properties are similar to those of larger population by estimating the ``unseen'' portion of the histogram. Two linear programming (LP) procedures are used to obtain this likely underlying histogram. The first LP algorithm finds the plausible histograms as follows. Since the finite data we obtained are the sampled ones from the unbounded population, the probability of drawing a topological type $i$ exactly $k$ times during $n$ independent trials follows the binomial distribution $B(n,p_{i})$, which can be approximated as a Poisson distribution ($P(np_{i},i)$). Hence, the first LP algorithm calculates the expected $i^{th}$ fingerprint entry and yields plausible histograms of which the fingerprints are the same as the expected fingerprint. The second LP algorithm selects the simplest distribution among the candidates based on Occam's razor. 
To validate the algorithm, we first apply the \textit{Unseen} estimator to ideal gas systems and compare $\hat{\mathcal{H}}$ to $\mathcal{H}$. We then build the following procedure to estimate the Voronoi entropy of a system based on the ideal gas results (see the Results and Discussion for the details). First, we perform five independent simulations for each condition and obtain 500 trajectories from each simulation. The number of molecules is 2,000. Second, we randomly select 300 trajectories of 2,500 configurations eight times and apply the \textit{Unseen} algorithm to each set of the trajectories. The estimated entropy data are averaged to obtain $\hat{\mathcal{H}}$.

  {Note that this algorithm does not reflect the spatial correlation between neighbors. Several measures have been suggested to reflect the spatial association \cite{boots2002local} to the Shannon entropy, but no algorithm has been proposed to consider both aspects. Later, we will see how this spatial correlation affects the results.}

\subsection{Thermodynamic excess entropy} The integration method of Deiters and Hoheisel \cite{hoheisel1979high} is used to calculate the thermodynamic excess entropy. In this method, the excess Gibbs free energy per particle ($G_{\rm{exc}}$) is calculated as:

\begin{equation}
    {\frac{G_{\rm{exc}}}{k_{B}T}=\int_{0}^{\rho}\frac{Z-1}{\rho}d\rho+Z-1}
    \label{eqn:deiters-hoheisel1}
\end{equation}
where $Z$ is the compressibility factor ($Z{\equiv}pV/RT$). Deiters-Hoheisel method constructs a function $Z(\rho)$ by fitting the compressibility factors obtained from a series of NVT simulations with a smoothing spline curve. $(Z-1)/\rho$ at the zero density converges to the second virial coefficient $B_{2}$, which is computed as:
\begin{equation}
    B_{2}=\lim_{\rho\rightarrow0}\frac{Z-1}{\rho}=-2\pi\int_{0}^{\infty}(e^{-\phi(r)/k_{B}T}-1)r^2{dr}
    \label{eqn:virial}
\end{equation}
where $\phi(r)$ is the interatomic potential. The equilibrium pressure and internal energy data are averaged every step during the production run ($5{,}000{,}000$ steps). The number of molecules is 2,000. Then, we use the trapezoidal rule to integrate the smoothing spline fitting function to evaluate the excess Gibbs energy per particle. The excess entropy is then obtained as:
\begin{equation}
    S_{\rm{exc}}=\frac{H_{\rm{exc}}-G_{\rm{exc}}}{k_{B}T}
\end{equation}
where $H_{\rm{exc}}$ is the excess enthalpy per particle, which is defined as $H_{\rm{exc}}=H-H_{\rm{ig}}$. The ideal gas enthalpy ($H_{\rm{ig}}$) is given as $H_{\rm{ig}}=(5/2)k_{B}T$.

\subsection{Calculation of transport properties}
The transport properties of the soft-sphere fluids are computed as follows. The diffusivity is estimated based on the vibrational density of states $\Psi$ defined as \cite{berens1983thermodynamics}:
\begin{equation}
    \Psi(\nu)=\frac{2}{k_{B}T}\sum_{j=1}^{N}\sum_{k=1}^{3}m_{j}\vec{\psi}_{j}^k(\nu)
    \label{eqn:density-of-states}
\end{equation}
where $\psi_{j}^{k}(\nu)$ is the spectral density of atom $j$ in the $k$ direction and $m_{j}$ is the mass of atom $j$. The spectral density is the square of the Fourier transform of the velocity.
\begin{equation}
    \vec{\psi}_{j}^{k}(\nu)=\lim_{\tau\rightarrow\infty} \left| \int_{-\tau}^{\tau} \vec{v}_{j}^{k}(t)e^{-i2\pi{\nu}t}dt \right|^2
    \label{eqn:spectral-density}
\end{equation}
Here, $\vec{v}_{j}^{k}(t)$ is the $k^{th}$ component of the velocity vector of the $j^{th}$ atom at time $t$. The diffusivity ($D$) of a system is then obtained from the intensity of $\Psi(\nu)$ at zero frequency as follows.
\begin{equation}
    D=\frac{\Psi(0)k_{B}T}{12mN}
    \label{eqn:diffusivity}
\end{equation}

\noindent {Note that decomposition of the spectral density into hard-sphere and harmonic oscillator contributions leads to another definition of the Frenkel line~\cite{yoon2018two}.}

The shear viscosity of a system is estimated by integrating the Green-Kubo integral \cite{green1954markoff,kubo1957statistical}.
\begin{equation}
    \eta=\frac{V}{k_{B}T}\int_{0}^{\infty}{\langle}{P}^{\alpha\beta}(t){\cdot}{P}^{\alpha\beta}(0)\rangle{dt}
    \label{eqn:viscosity}
\end{equation}
where ${P}^{\alpha\beta}(t)$ ($\alpha$, $\beta=x,y,$ and $z$) is the off-diagonal elements of the pressure tensor, which is given as:
\begin{equation}
    {P}^{\alpha\beta}=\sum_{j=1}^{N}\frac{m_{j}{v}_{j}^{\alpha}{v}_{j}^{\beta}}{V}+\frac{\sum_{j}^{N'}{r}_{j}^{\alpha}{{f}_{j}^{\beta}}}{V}
    \label{eqn:stress-tensor}
\end{equation}
Here, ${r}_{j}^{\alpha}$ is the $\alpha^{th}$ component of the position vector ${r}$ of the $j^{th}$ atom and ${f}_{j}^{\beta}$ is the $\beta^{th}$ component of the force vector $\vec{f}$ exerted on the $j^{th}$ atom. 

Unfortunately, calculating $\eta$ using the Green-Kubo integral is difficult due to the low signal-to-noise ratio; the stress autocorrelation function given in Eqn.~(\ref{eqn:viscosity}) does not smoothly converge to zero. Hence, we combine the methods proposed by Nevins \cite{nevins2007accurate} and Zhang \cite{zhang2015reliable} as follows. We perform ten independent NVT simulations with different initial configurations and initial velocities for each thermodynamic condition. The timestep is equal to the equillibration run (2 fs), and the simulation duration is 2 ns. The stress autocorrelation functions obtained from independent simulations are averaged and truncated at the cutoff time ($t_{\rm{cut}}$) at which the absolute magnitude of the stress autocorrelation function decreases under its initial value by a factor of $10^{-3}$. Then, a two-term exponential function is fitted to the truncated stress autocorrelation function. The viscosity is calculated by integrating the fitted stress autocorrelation function.

\section{Results and Discussion}
\subsection{Voronoi entropy of the ideal gas}
\begin{figure}
    \centering
    \includegraphics{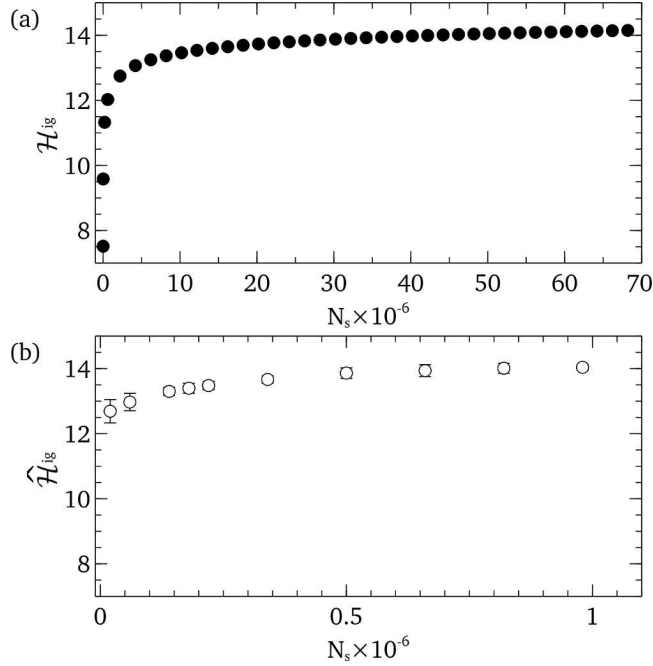}
    \caption{Shannon entropy of ideal gas system estimated from (a) the observed probabilities ($\mathcal{H}$) and (b) the \textit{Unseen} algorithm ($\hat{\mathcal{H}}$). $\mathcal{H}$ does not vary significantly ($\mathcal{H}\sim14.04$) when the sample size ($N_{s}$) is larger than $430{,}000{,}000$, whereas $\hat{\mathcal{H}}$ becomes close to 14.00 when $N_{s}$ is larger than $5{,}000{,}000$.}
    \label{fig:ig_shannon}
\end{figure}
We first estimate the Voronoi entropy of ideal gas to determine the number of samples and trials for the soft-sphere models. Fig.~\ref{fig:ig_shannon} compares $\hat{\mathcal{H}}$ and $\mathcal{H}$ of the ideal gas. The ideal gas configurations are generated by distributing $N=2{,}000$ points in a cubic box randomly. As shown in Fig.~\ref{fig:ig_shannon}a, $\mathcal{H}_{\rm{ig}}$ slowly increases as the sampled number of configurations (molecules) increases. When the sample size ($N_{s}$) is larger than $30{,}000{,}000$, $\mathcal{H}_{\rm{ig}}$ does not vary significantly. On the contrary, the Voronoi entropy estimated from the \textit{Unseen} becomes similar to $\hat{\mathcal{H}}_{\rm{ig}}\sim14.00$ when the sample size is larger than $1{,}000{,}000$ (Fig.~\ref{fig:ig_shannon}b). The order of the magnitude of the sample size to obtain $\hat{\mathcal{H}}_{\rm{ig}}$ similar to $\mathcal{H}_{\rm{ig}}$ agrees with that proposed by Valiant and Valiant ($30{,}000{,}000/\log(30{,}000{,}000){\sim}\mathcal{O}(10^6)$). Since the population of the topological types of the ideal gas system is larger than those of any other system, $\mathcal{O}(10^2)$ trajectories of $N=2{,}000$ molecules are enough to estimate the Voronoi entropy at all thermodynamic conditions studied in this work. {The estimated Voronoi entropy of the ideal gas ($\hat{\mathcal{H}}_{ig}\sim14.04$) can be used to define the Voronoi excess entropy, which is given as $\hat{\mathcal{H}}_{exc}\equiv\hat{\mathcal{H}}-\hat{\mathcal{H}}_{ig}$.}

\subsection{Voronoi entropy of repulsive $n$-6 fluids}
\begin{figure}
    \centering
    \includegraphics{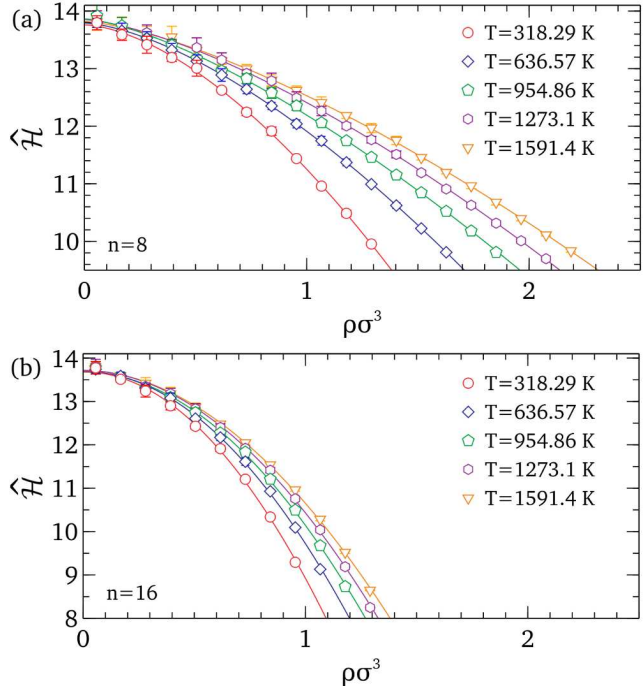}
    \caption{Dependence of $\hat{\mathcal{H}}$ on the bulk density ($\rho$). (a) $n=8$ and (b) $n=16$. It shows a power-law dependence on the bulk density. {The $\hat{\mathcal{H}}$, the upper bound of the Voronoi entropy, is calculated from the Unseen algorithm.}}
    \label{fig:shannon-entropy-ssp}
\end{figure}
Fig.~\ref{fig:shannon-entropy-ssp} shows $\hat{\mathcal{H}}$ of the fluids modeled with the repulsive $8-6$ potential and $16-6$ potential. $\hat{\mathcal{H}}$ slowly decreases in the low-density region, but ${\lvert}d\hat{\mathcal{H}}/d\rho{\rvert}$ increases as the density increases; it shows the power-law dependence on the density (Eqn.~(\ref{eqn:power-law-shannon})).
\begin{equation}
    \hat{\mathcal{H}}=a\rho^b+c
    \label{eqn:power-law-shannon}
\end{equation}
where $a$, $b$, and $c$ are fitting parameters. The power-law equations fitted to different isotherms converge to $\hat{\mathcal{H}}\sim14.0$ as the density decreases, which agrees with the Voronoi entropy of ideal gas. The power-law equation indicates that the Voronoi entropy of a system decreases as the density approaches the freezing density. The decrease of the Voronoi entropy can be understood based on the free volume theory \cite{nezhad2017estimation,yoon2018topological}. As the system density increases, the distances between neighbor atoms surrounding the central atom become shorter. When they are so close that they hinder each others' diffusive motions, the number of ways to place neighbors around a central atom without an increase of the potential-energy decreases. As a result, the Voronoi entropy drastically decreases when the bulk density is high. A large discrepancy between $\mathcal{H}$ and $\hat{\mathcal{H}}$ in the low-density region reflects this phenomenon. Since the set of possible topological types are drastically large and cannot be sufficiently sampled in low-density systems, $\mathcal{H}$ of the low-density fluid is much lower than $\hat{\mathcal{H}}$, whereas that of the high-density fluid is similar to $\hat{\mathcal{H}}$.
\begin{figure}
    \centering
    \includegraphics{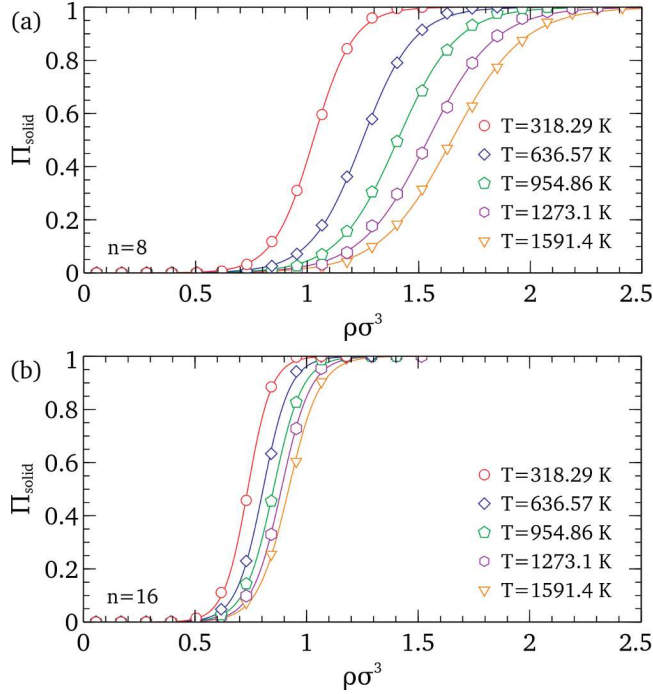}
    \caption{Dependence of the solid-like fraction ($\Pi_{\rm{solid}}$) on the bulk density ($\rho$). {For the definition of the solid-like and gas-like states, see Sec. \ref{sec:fl}.} (a) $n=8$ and (b) $n=16$. It shows a sigmoidal dependence on the bulk density for all temperatures and repulsive exponents.}
    \label{fig:shannon-entropy-ssp2}
\end{figure}
\begin{figure}
    \centering
    \includegraphics{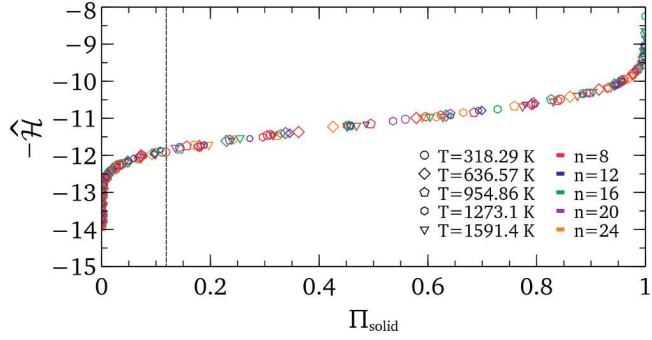}
    \caption{A relation between the Voronoi entropy ($\hat{\mathcal{H}}$) and the fraction of solid-like molecules ($\Pi_{\rm{solid}}$). $\hat{\mathcal{H}}$ has a one-to-one correspondence with $\Pi_{\rm{solid}}$. This correspondence relation enables to redefine the Frenkel line as a set of isomorphic states, which demarcate the fluid region into the non-rigid and the rigid regions.}
    \label{fig:shannon-pi}
\end{figure}

Fig.~\ref{fig:shannon-entropy-ssp2} shows $\Pi_{\rm{solid}}$ of the repulsive 8-6 fluids {(For the definition of $\Pi_{\rm{solid}}$, see Sec. \ref{sec:fl}.)}. As shown in our earlier works \cite{yoon2018topological,yoon2019topological}, it starts to steeply increase near the dynamic crossover densities and reaches unity near the freezing densities at constant temperatures. The dependence of $\Pi_{\rm{solid}}$ on the bulk density is well expressed by the sigmoidal function for all conditions based on the theory of fluid polyamorphism \cite{anisimov2018thermodynamics}.
\begin{equation}
    \Pi_{\rm{solid}}=\frac{1}{1+a\exp(b\rho)}
\end{equation}
Both the Voronoi entropy and the fraction of solid-like molecules at different isotherms become close to each other when the repulsive exponent increases. They become close to each other and ultimately collapse to a single line when the repulsive exponent ($n$) is infinite \cite{yoon2019topological}. 

Since both Voronoi entropy and the fraction of solid-like molecules are defined from the topological types of the Voronoi cells, it can be expected that both parameters are deeply related. Fig.~\ref{fig:shannon-pi} shows that a one-to-one correspondence relation exists between $\hat{\mathcal{H}}$ and $\Pi_{\rm{solid}}$. This one-to-one correspondence relation has two implications. First, it substantiates that the topological framework captures the isomorphism observed along the freezing line. Since $\Pi_{\rm{solid}}$ reaches unity at the freezing densities, $\hat{\mathcal{H}}$ becomes approximately $\hat{\mathcal{H}}\sim9.0$ along the freezing line. Second, the Voronoi entropy is a linear function of $\Pi_{\rm{solid}}$ over the interval of $10\leq\hat{\mathcal{H}}\leq12$. Considering that the Frenkel line is an iso-$\Pi_{\rm{solid}}$ line ($\Pi_{\rm{solid}}=0.1159$) \cite{yoon2018topological}, this result implies that the Frenkel line can be related to the qualitative different regimes of the collective particle dynamics as we expected. This expectation is further discussed in the next section.   

\subsection{Isomorph theory based on the Voronoi entropy}
\begin{figure}
    \centering
    \includegraphics{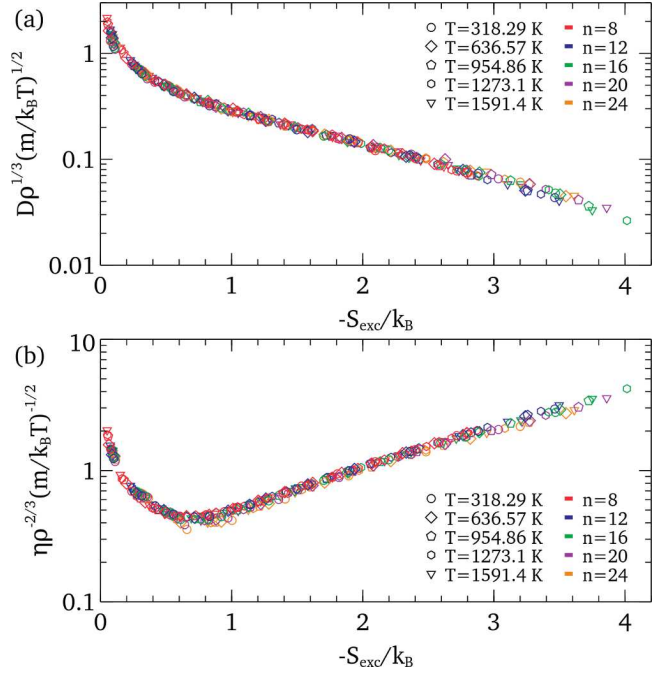}
    \caption{Thermodynamic excess entropy scaling of (a) diffusivity ($\tilde{D}$) and (b) shear viscosity ($\tilde{\eta}$). For all repulsive $n-6$ fluids, the Rosenfeld diffusivity and shear viscosity collapse to single lines.}
    \label{fig:thermo-scaling}
\end{figure}

  {We test the validity of the Rosenfeld scaling law for repulsive $n$-6 fluids.} Fig.~\ref{fig:thermo-scaling} shows that the Rosenfeld diffusivity ($\tilde{D}$) and viscosity ($\tilde{\eta}$) of all simple fluids modeled with repulsive $n$-6 potential collapse to single lines {as the isomorph theory for R-simple fluids predicts}. They are defined as:
\begin{equation}
    {\tilde{D}=D\rho^{1/3}\left(\frac{m}{k_{B}T}\right)^{1/2};\quad\tilde{\eta}=\eta\rho^{-2/3}\left(\frac{m}{k_{B}T}\right)^{-1/2}}
\end{equation}
Both collapsing lines show a similar dependence on $-S_{\rm{exc}}$ observed by previous works \cite{rosenfeld1977relation,ding2015equilibrium,chopra2010use,jakse2016excess}. $\tilde{D}$ curve steeply decreases as $-S_{\rm{exc}}$ increases to a certain extent. It shows an exponential dependence on $-S_{\rm{exc}}$ when $S_{\rm{exc}}$ decreases. $\tilde{\eta}$ shows more complex dependence on $-S_{\rm{exc}}$. It decreases to its minimum in the middle-density region, and increases as $-S_{\rm{exc}}$ increases.

\begin{figure}
    \centering
    \includegraphics{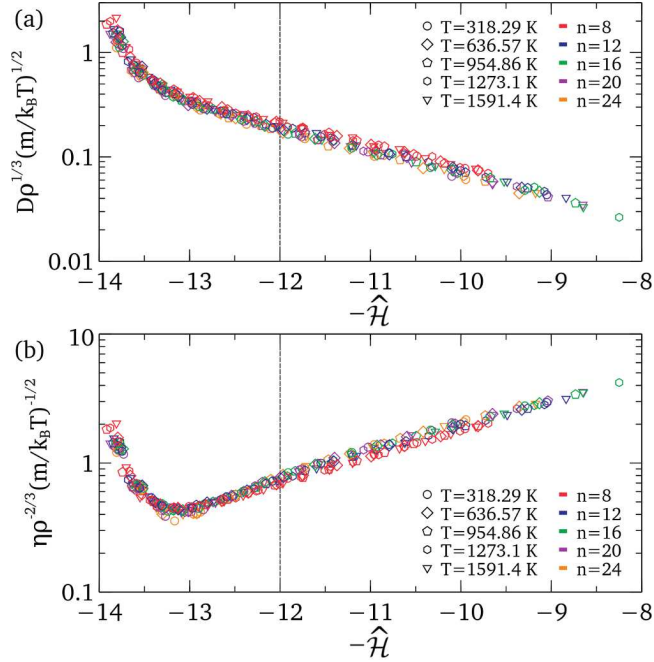}
    \caption{Voronoi entropy scaling of (a) diffusivity ($\tilde{D}$) and (b) shear viscosity ($\tilde{\eta}$) of fluids modeled with repulsive $n$-6 potentials. The dotted lines denote the dynamic crossover Voronoi entropies ($\hat{\mathcal{H}}=\hat{\mathcal{H}}_{cr}$), which are estimated based on the $\Pi_{\rm{solid}}$-$\hat{\mathcal{H}}$ curve.}
    \label{fig:shannon-scaling}
\end{figure}

\begin{table}
    \centering
    \begin{tabular}{cccc}
    \hline
    n & Slope & Intersect & $R^2$\\
    \hline
    8 & 1.479 & -0.013 & 0.996 \\
    12 & 1.438 & -0.043 & 0.996 \\
    16 & 1.392 & -0.051 & 0.994 \\
    20 & 1.337 & -0.038 & 0.996 \\
    24 & 1.286 & -0.009 & 0.996 \\
    \hline
    \end{tabular}
    \caption{Fitting parameters for the linear relation between the thermodynamic excess entropy and the Voronoi excess entropy ($-\hat{\mathcal{H}}_{exc}=-pS_{exc}/k_{B}+q$). Regardless of the repulsive exponents, the coefficient of determination ($R^2$) are higher than $0.99$, and the intersects ($q$) are close to zero.}
    \label{tab:thermo-info}
\end{table}
Fig.~\ref{fig:shannon-scaling} shows the dependence of $\tilde{D}$ and $\tilde{\eta}$ on the Voronoi entropy. Similar to the thermodynamic excess entropy, they collapse to their own single lines to a good approximation. In addition, the shapes of the collapsing lines are similar to those obtained when $-S_{\rm{exc}}/k_{B}$ is used to scale the transport properties. Simultaneously, they substantiate that the Frenkel line is a set of isomorphic states ($\hat{\mathcal{H}}\sim12.0$) where the collective particle dynamics qualitatively changes. When $\hat{\mathcal{H}}$ is greater than the topological crossover diversity, $\tilde{D}$ shows a power-law dependence on $\hat{\mathcal{H}}$. As the density increases, $\hat{\mathcal{H}}$ drastically decreases and exponential decay of $\tilde{D}$ is observed. $\tilde{\eta}$ also shows an exponential dependence on $-\hat{\mathcal{H}}$ when $\hat{\mathcal{H}}$ is lower than the topological crossover diversity. All these results support that the Frenkel line, a rigid-nonrigid transition line, is a good candidate that demarcates the fluid region considering the collective particle dynamics.

\begin{figure}
    \centering
    \includegraphics{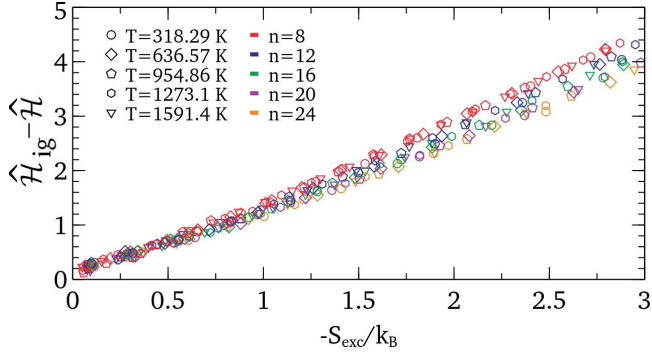}
    \caption{A linear relation between the thermodynamic excess entropy and the Voronoi {excess} entropy. {Although all models show the linear relationship, $\hat{\mathcal{H}}_{exc}(S_{\rm{exc}})$ curves of repulsive $n-6$ fluids are not consistent with each other. This discrepancy would bring about by ignoring the spatial correlation between neighbors.}}
    \label{fig:thermoinfo}
\end{figure}

Meanwhile, it should be noted that the scaling results from $\hat{\mathcal{H}}$ show a slight but systematic discrepancy compared to those from $-S_{\rm{exc}}$. For the same repulsive $n-6$ models, both $\tilde{D}$ and $\tilde{\eta}$ curves from different temperatures agree with each other. In contrast, the extent of the data collapse is low for different repulsive exponents compared to the thermodynamic excess entropy scaling result. 

Fig.~\ref{fig:thermoinfo} shows the parity plot of {$-S_{\rm{exc}}$} and {$-\hat{\mathcal{H}}_{exc}$}. {As the repulsive exponent increases, the slope of $-\hat{\mathcal{H}}_{exc}(-S_{exc})$ decreases. This small discrepancy between slopes of different models would occur because the algorithm ignores the mutual dependence between the topological types of neighbors. At the same thermodynamic excess entropy, the bulk density of a system increases as the repulsive exponent ($n$) decreases. Bearing in mind that the cutoff radius increases as $n$ decreases, the spatial correlation between neighbors will be high when $n$ is low.} 

  {Despite these limitations, $-S_{\rm{exc}}$ and $-\hat{\mathcal{H}}_{exc}$ have a linear relation, and the intersects of the fitted equations are close to zero (Table \ref{tab:thermo-info}).} This linearity indicates that the influence of the spatial correlation is little compared to the total entropy contribution. Hence, we can see that the classical idea about the equivalence of the Shannon entropy and the thermodynamic entropy holds \cite{pfleger2015discrete}. 

\section{Conclusion}
This work demonstrates that the isomorph theory can be extended to the molecular level in conjunction with the topological framework and the information theory. In this framework, two systems are regarded to be isomorphic if their topological diversities (Voronoi entropies) are equal. The Voronoi entropy of the Bernoulli distribution, in which the number of categories is infinite, can be estimated based on the fingerprint of the distribution. Similar to the thermodynamic excess entropy, the Voronoi entropy can work as a scaling parameter to collapse the transport properties of soft-sphere fluids. The Voronoi entropy scaling results are satisfactory but also show slight but noteworthy deviations compared to the thermodynamic excess entropy scaling. These systematic deviations come from the limit of the {proposed method; it ignores the entropic contribution of the particles which are not nearest neighbors but influence the net force exerted on the central particle.} Lastly, a qualitatively different dependence of the transport properties on the $\hat{\mathcal{H}}$ and $S_{\rm{exc}}$ can be understood based on the rigid-nonrigid dynamic crossover across the Frenkel line. {Since the isomorph theory is quasi-universal for various types of potentials, it would be interesting to understand the quasi-universal characteristics of broader classes of fluid models based on the topological framework in conjunction with the information theory.} 

\section{Supplementary material}
Supplementary material includes numerical data that can help understand and reproduce the results obtained in this work. 

\section{Acknowledgements}
M.Y.H. and W.B.L. acknowledge the support by Creative Materials Discovery Program through the National Research
Foundation of Korea (NRF) funded by Ministry of Science and ICT (2018M3D1A1058624).  E.A.L. gratefully acknowledges the generous support of the US National Science Foundation through Award DMR-1507013.

\end{document}